\documentclass[prl,twocolumn,superscriptaddress,preprintnumbers,nofootinbib,nopacs]{revtex4-1}

\usepackage{amsmath}
\usepackage{amssymb}
\usepackage{amsbsy}
\usepackage{amsfonts}
\usepackage{graphicx}
\usepackage{hyperref}
\usepackage{bibunits}
\usepackage{comment}
\usepackage{resizegather}
\usepackage{xr}
\newcommand{\bfig}{\begin{center}\vskip 0.5em}
\newcommand{\efig}{\end{center}\vskip 0.5em}
\newcommand{\eqnref}[1]{Eq.~(\ref{#1})}

\newcommand{\cV}{U'}

\newcommand{\ket}[1]{|#1\rangle}
\newcommand{\bra}[1]{\langle#1|}
\newcommand{\Uf}{U_{\mathrm{f}}}

\usepackage{color}

\begin{document}

%%%%%%%%%%%%

\title{Floquet Time Crystals}
%\title{Floquet Time Crystals}

\author{Dominic V. Else}
\affiliation{Physics Department, University of California,  Santa Barbara, California 93106, USA}

\author{Bela Bauer}
\affiliation{Station Q, Microsoft Research, Santa Barbara, California 93106-6105, USA}

\author{Chetan Nayak}
\affiliation{Station Q, Microsoft Research, Santa Barbara, California 93106-6105, USA}
\affiliation{Physics Department, University of California,  Santa Barbara, California 93106, USA}

\begin{abstract}
We define what it means for time translation symmetry to be spontaneously broken
in a quantum system, and
show with analytical arguments and numerical simulations that this occurs in
a large class of many-body-localized driven systems with discrete time-translation symmetry.
%Our construction builds on recent advances in the study of Floquet many-body localized
%systems. Our definitions enable the identification of time-translational symmetry breaking
%phases -- time crystals -- in numerical simulations that we carry out and experiments that we propose.
\end{abstract}

\maketitle

\begin{bibunit}[apsrevmod]

\paragraph{Introduction.}
Spontaneous symmetry-breaking (SSB) is a pivotal concept in physics, with implications
for condensed matter and high-energy physics. It occurs when
the ground state or low-temperature states of a system fail to be invariant under symmetries
of the Hamiltonian.
The Ising model is a prototypical example for this behavior: Here, the symmetry is a
simultaneous flip of all the spins, which leaves the energy of a state unchanged.
In the ferromagnetic phase, low-energy states are formed with a non-zero magnetization.
\begin{comment}
and thus the action of the symmetry on such a state creates a new
low-energy state with equal and opposite magnetization.
For finite systems, the eigenstates are even and odd superpositions of such states,
which become exactly degenerate in the thermodynamic limit.
Such superpositions -- often referred to as cat states --  cannot be distinguished by any local measurement,
cannot be prepared in an experiment and are unstable to decoherence.
A weak, even infinitesimal, field will disrupt the superposition and
drive the system into a state in which the spins point
preferentially in the direction of the field. SSB can be defined either as
the development of a finite symmetry-breaking magnetization in response
to an infinitesimal field or, alternatively, as the condition that low-lying energy eigenstates that
are also eigenstates of the symmetry must be cat states.
\end{comment}
%
For almost every symmetry imaginable, there is a model whose ground state breaks it:
crystals break the continuous translational and rotational symmetries of Coulomb interactions;
magnetically ordered materials break time-reversal symmetry and spin symmetry, and superfluids break global gauge symmetry.
The lone holdout, thus far,
has been time-translation symmetry. In this paper, we give a definition of
time-translation symmetry breaking, and construct an example of this behavior in
a driven many-body localized system.

\paragraph{Definition of Time Translation Symmetry-Breaking.}
Systems that spontaneously break time-translation symmetry (TTS) have been dubbed ``time crystals,'' in analogy
with ordinary crystals, which break spatial translational
symmetries~\cite{Wilczek12,Shapere12}. Even defining this notion correctly
requires considerable care, and putative models have proven inconsistent
\cite{Li13,Bruno13a,Bruno13b,Bruno2013,Nozieres13,Volovik2013,Sacha16}.
%However, this concept
%has been controversial due to the absence of an appropriate definition and a canonical model that satisfies it~\cite{Li13,Bruno13a,Bruno13b,Bruno2013,Nozieres13,Volovik2013,Sacha16}.
The most obvious definition of time-translation symmetry breaking (TTSB) would
be that the expectation values of observables are time-dependent in thermal
equilibrium. However, this is clearly impossible, since a thermal equilibrium
state $\rho = \frac{1}{Z}e^{-\beta H}$ is time-independent by construction
(because $[\rho,H] = 0$). A more sophisticated definition of TTSB in terms of
correlation functions in the state $\rho$ has been proposed -- and ruled out
by a no-go theorem -- in Ref.~\cite{Watanabe15}.

Therefore, we must look beyond strict thermal equilibrium. This should not be
too surprising, as the state $\rho$ preserves \emph{all} the symmetries of $H$, which would suggest that \emph{no} symmetry can be spontaneously broken. For symmetries other than time translation, the resolution to this paradox is well-known: in a system with a spontaneously broken symmetry, there is ergodicity-breaking and the lifetime of a symmetry-breaking state diverges as the system size grows. Thus, in the thermodynamic limit, the state $\rho$ is unphysical and is never reached. This suggests that an analogous phenomenon should be possible for time translation symmetry, where the time taken to reach a time-independent steady state (such as the thermal state $\rho$) diverges exponentially with system size.

To turn these considerations into a more useful definition, we observe that, in a quantum system, the ergodicity-breaking in a phase with a spontaneously broken symmetry can be seen at the level of eigenstates. For example, the \emph{symmetry-respecting} ground states of an Ising ferromagnet are $\ket{\pm} = \frac{1}{\sqrt{2}}(\ket{\uparrow \cdots \uparrow} \pm \ket{\downarrow \cdots \downarrow}$. Such long-range correlated ``cat states'' are unphysical, will immediately decohere given any coupling to the environment, and can never be reached in finite time by any unitary time evolution starting from a short-range correlated starting state. On the other hand, the ``physical'' combinations $\ket{\uparrow \cdots \uparrow}$ and $\ket{\downarrow \cdots \downarrow}$ break the Ising symmetry.

In TTSB case, we also need to invoke the intuition that oscillation under time evolution
requires the superposition of states whose phases wind at different rates. That is, whereas in the Ising ferromagnet the two cat states $\ket{\pm}$ are degenerate in the thermodynamic limit, in a time-crystal they would need to have \emph{different} eigenvalues under the time-evolution operator. Indeed, consider for simplicity a discrete time evolution operator $\Uf$ (which describes periodically driven ``Floquet'' systems as we discuss further below.) Suppose that that the states $\ket{\pm}$ have eigenvalues $e^{i \omega_{\pm}}$ under $\Uf$. Then, although the unphysical cat states $\ket{\pm}$ are time-invariant (up to a phase), a \emph{physical} state such as $\ket{\uparrow \cdots \uparrow}$ will evolve according to $(\Uf)^n \ket{\uparrow} \propto \cos(\omega n) \ket{\uparrow \cdots \uparrow} + \sin(\omega n) \ket{\downarrow \cdots \downarrow}$, where $\omega = (\omega_{+} - \omega_{-})/2$.

The above considerations motivate two equivalent definitions of TTSB, using
the following terminology/notation.
We will say that a state $|\psi\rangle$ has {\it short-ranged
correlations} if, for any local operator $\Phi(x)$,
$\langle\psi| \Phi(x) \Phi(x')|\psi\rangle - \langle\psi| \Phi(x)|\psi\rangle \langle\psi|\Phi(x')|\psi\rangle \rightarrow 0$ as $|x-x'|\rightarrow \infty$,
i.e. if cluster decomposition holds.
Note that the superpositions defined above are not short-range correlated under this definition, while a state
such as $|\uparrow \uparrow \ldots \uparrow \rangle$ is. We assume that time-evolution is described by a time-dependent Hamiltonian $H(t)$, with a discrete time translation symmetry such that $H(t + T)$ for some $T$. Note that we have not assumed a \emph{continuous} time translation symmetry, which
will allow us to consider ``Floquet'' systems driven at a frequency $\Omega = 2\pi/T$.
Let $U(t_1,t_2)$ be the corresponding time evolution operator from time $t_1$ to $t_2$.
We now define (in the thermodynamic limit):

{\bf TTSB-1}: TTSB occurs if for each $t_1$, and for every state $|\psi(t_1)\rangle$ with short-ranged correlations,
there exists an operator $\Phi$ such that $\bra{\psi(t_1+T)} \Phi \ket{\psi(t_1+T)} \neq \bra{\psi(t_1)} \Phi \ket{\psi(t_1)}$, where $\ket{\psi(t_1 + T)} = U(t_1 + T, t_1) \ket{\psi(t_1)}$.

{\bf TTSB-2}:
TTSB occurs if the eigenstates of the Floquet operator $\Uf \equiv U(T,0)$ cannot be short-range correlated.

%Note that neither of these definitions are strictly ground state properties or properties of individual eigenstates which, as discussed above, cannot show
%non-trivial time dependence; they are properties of a time-translation operator.
 In what follows, we will show how to construct a time-dependent Hamiltonian $H(t)$ which satisfies the conditions for TTSB given above. In such a system,
even though the time-evolution
is invariant under the discrete TTS generated by time translation by $T$, the expectation value of some observables is only invariant under translations by
$nT$ for some $n>1$. In other words, the system responds at a \emph{fraction} $\Omega/n$ of the original driving frequency.

The first definition puts the time-dependence front and center and is directly connected to how
TTSB would be observed experimentally: prepare a system in a short-range correlated state and observe its subsequent time-evolution, which will not be invariant under the TTS of the time evolution operator. But since, in a Floquet eigenstate,
observables would necessarily be invariant under the discrete TTS generated by time translation by $T$, definition TTSB-1 implies that
Floquet eigenstates cannot be short-range correlated, thereby implying TTSB-2. Conversely, if it is impossible to find Floquet eigenstates that are
short-range correlated (which is TTSB-2), then it means that short-range correlated states can only be formed by taking superpositions of
Floquet eigenstates with different eigenvalues. In such states, observables will not be invariant under the discrete TTS generated by time translation by $T$,
thereby implying TTSB-1. Hence, the two definitions are equivalent. The second definition will prove to be particularly useful
for analyzing the results of
numerical exact diagonalization of the Floquet operator. When discrete TTS by $T$ is broken down to TTS by $nT$, the eigenstates
of $\Uf$ must be superpositions of $n$ different short-range-ordered states.\footnote{To see this, note that we can choose a basis of short-range correlated eigenstates for $(\Uf)^n$. By assumption, such states cannot be eigenstates of $(\Uf)^k$ for $0 < k < n$. Therefore, $\Uf$ generates an orbit of $n$ \emph{different} short-range correlated states. An eigenstate of $\Uf$ is an equal-weight superposition over such an orbit.} Then, in any Floquet eigenstate, the mutual information $I(A,B) \equiv S_A + S_B - S_{AB}$, where $A$ and $B$ are spatially separated regions of the system and $S_X$ is the von Neumann entropy of the reduced density matrix for region $X$, satisfies $I(A,B) \rightarrow \ln n$ as the system size as well as the sizes of the regions $A$ and $B$ and their separation is taken to infinity \cite{Jian15}.

\paragraph{Floquet-Many-Body-Localization.}
Generic translationally invariant many-body Floquet systems likely cannot have TTSB, as their eigenstates
resemble infinite temperature states and hence are short-range correlated~\cite{DAlessio2014,Lazarides2014,Ponte2015a}.\footnote{Nevertheless, an initial state that is not an eigenstate could potentially heat very slowly, leading to non-trivial intermediate-time dynamics~\cite{Abanin2015,Abanin2015b,Kuwahara2015,Mori2015,Chandran2015}.}
This is analogous to the fact (which follows from the results of
Ref.~\cite{Watanabe15}) that for continuous time-translation symmetry, TTSB is
impossible so long as the eigenstate thermalization hypothesis
(ETH)\cite{Deutsch1991,Srednicki1994,Rigol2008,DAlessio2015} is
satisfied.
However, we can build upon recent developments in the study of
Floquet-many-body-localized (Floquet-MBL)
systems~\cite{Abanin2014,Ponte15a,Ponte15b,Lazarides15,Iadecola15,Khemani15b,vonKeyserlingk16a,Else16a,Potter16,Roy16,vonKeyserlingk16b},
for which the eigenstates do not resemble infinite temperature states.
Instead, the Floquet states of such systems exhibit the characteristics of the energy eigenstates of
static MBL~\cite{Basko06a,Basko06b,Oganesyan07,Znidaric08,Pal10,Bardarson12,Huse2013a,Serbyn13a,Serbyn13b,Huse14} systems:
the eigenstates are local product states, up to finite-depth unitary quantum circuits~\cite{Bauer13}.

In MBL systems, all eigenstates (of the Hamiltonian in the static case or of the Floquet operator in the driven case) behave as ground states
and, therefore, SSB or topological order can occur in all eigenstates~\cite{Huse13,Bauer13,Bahri15}. In the SSB case, simultaneous eigenstates of
the Floquet operator and of the Cartan subalgebra of the symmetry generators cannot be short range correlated.
TTSB-2 can then be viewed as a special case of this in which there are no other symmetry generators besides $\Uf$.

In the next paragraph, we construct a Floquet operator and show that it exhibits discrete TTSB. In subsequent paragraphs, we 
show that this soluble Floquet operator sits in a finite window in parameter space over which TTSB occurs -- i.e. that there is
a TTSB {\it phase}.
Models which exhibit TTSB (though not identified as such) have previously been
considered in Refs.~\cite{Khemani15b,vonKeyserlingk16b}. These models also break
another symmetry spontaneously, but this is not essential to achieve TTSB. Our
model will be a generalization of that of
Refs.~\cite{Khemani15b,vonKeyserlingk16b}, with the extra symmetry explicitly broken.

\paragraph{Model and Soluble Point.}
We consider one-dimensional spin-$1/2$ systems with Floquet unitaries of the form:
\begin{equation}
\label{eqn:soluble-U_f}
\Uf =  \exp\left(-i {t_0} H_\text{MBL}\right) \, \exp\left(i \frac{\pi}{2} {\sum_i} {\sigma^x_i}\right)
\end{equation}
In this stroboscopic time evolution, the Hamiltonian ${\sum_i} {\sigma^x_i}$ is applied for a time interval $\pi/2$, which
has the effect of flipping all of the spins since $\exp(i \frac{\pi}{2} {\sum_i} {\sigma^x_i}) = {\prod_i} i{\sigma^x_i}$. This is followed
by time evolution for an interval $t_0$ under the Hamiltonian
\begin{equation}
\label{eqn:H_MBL}
H_\text{MBL} = \sum_i \left(J_i \sigma^z_i \sigma^z_{i+1} + h^z_i \sigma^z_i + h^x_i \sigma^x_i \right)
\end{equation}
where $J_i$, $h^z_i$, and $h^x_i$ are uniformly chosen from $J_i \in [\frac{1}{2},\frac{3}{2}]$,
$h^z_i \in [0,1]$, $h^x_i \in [0,h]$ where small $h$ is the regime of interest.
The period of the drive is $T = {t_0} + \frac{\pi}{2}$.
For $h=0$, the eigenstates of $H_\text{MBL}$
are eigenstates of the individual $\sigma^z_i$. Call such an eigenstate $|\{s_i \}\rangle$
with $s_i=\pm 1$ so that $\sigma^z_k |\{s_i \}\rangle = s_k |\{s_i \}\rangle$. Then $H|\{s_i \}\rangle = ({E^+}(\{s_i \}) + {E^-}(\{s_i \}) )|\{s_i \}\rangle$ where ${E^+}(\{s_i \}) = \sum_i (J_i s_i  s_{i+1}) $ and $ {E^-}(\{s_i \}) = \sum_i (h^z_i s_i)$.
The Floquet eigenstates are $(\exp(i t_0 {E^-}(\{s_i \}) /2) |\{s_i \}\rangle \pm \exp(-i t_0 {E^-}(\{s_i \}) /2)|\{-s_i \}\rangle)/\sqrt{2}$,
and the corresponding Floquet eigenvalues are $\pm \exp(i t_0 {E^+}(\{s_i \}) )$. Hence, TTSB-2 is satisfied for $h=0$.

\paragraph{Stability of TTSB.}
We now argue that the preceding conclusions are no fluke: weak perturbations of the Floquet operator, such as non-zero $h$
or deviations of the length of the second time-interval from $\frac{\pi}{2}$, do not destroy TTSB, so long as a reasonable but non-trivial
assumption about resonances holds.
Ordinarily, there would be little doubt that SSB of a discrete symmetry is stable to weak perturbations at zero-temperature
in 1D. But since the symmetry in question is TTS, more care seems necessary.

To build confidence in the stability of TTSB, we can exploit the discrete local connectivity of fully MBL systems: that is, for any eigenstate $\ket{i}$, and point $x$, there is only a finite number of eigenstates $\ket{j}$ such that the matrix elements $\bra{i} \Phi(x) \ket{j} \neq 0$ for some operator $\Phi(x)$ acting locally at $x$. In particular, generically the (quasi-)energy difference $\omega_j - \omega_i$ for eigenstates connected in this way will not be close to zero. In systems with such a \emph{local spectral gap}, one expects that \emph{local perturbations perturb locally}~\cite{Hastings2005,Bravyi2010,Bravyi2011,Bachmann2012}, or more precisely, that there exists a single local unitary $\mathcal{U}$ (that is, a unitary which can be expressed as the time evolution of a local Hamiltonian $S$) which relates perturbed eigenstates to unperturbed eigenstates~\cite{Bauer13}. Such a local unitary $\mathcal{U}$ cannot possibly connect short-range correlated states with the long-range correlated eigenstates found above. Therefore, the eigenstates of the perturbed Floquet operator still satisfy TTSB-2.

We make these ideas more precise in the Supplementary Material. There, we construct the unitary $\mathcal{U}$ order-by-order in perturbation theory and show that it remains local at all orders, provided that the local spectral gap condition holds. The skeptic might argue, however, that there will always be rare regions (known as ``resonances'') in which the local spectral gap is arbitrarily small, and that this will spoil the convergence of the perturbation theory. A rigorous treatment of resonances is a difficult problem; however, the principle of ``local perturbations perturb locally'' has in fact been proven (given certain reasonable assumptions), at least for a particular model of stationary MBL~\cite{Imbrie2014a}.

\begin{figure}
  \includegraphics{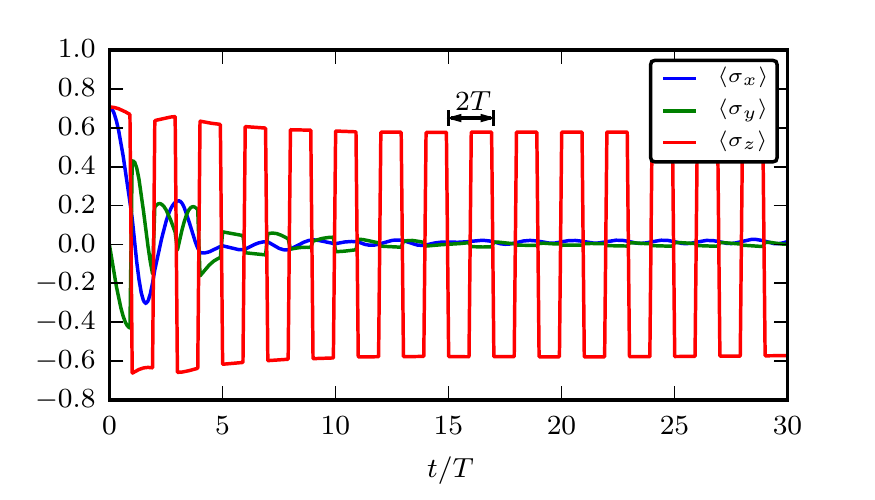}
  \includegraphics{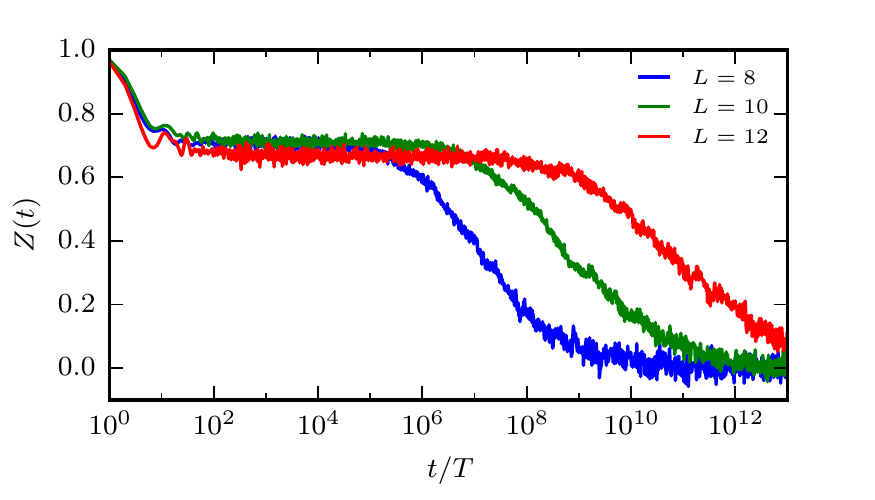}
  \caption{(Color online)
     The time evolution of a short-range correlated initial state satisfies TTSB-1 for $h=0.3$.
	Top Panel: the time-dependence of
	the disorder-averaged $\langle \sigma^x_i \rangle$, $\langle \sigma^y_i \rangle$, and $\langle \sigma^z_i \rangle$
	show that the former two decay rapidly while the latter displays persistent
	oscillations. (The spin-flip part of the Floquet operator is here taken to be
	applied instantaneously.)
	Bottom Panel: The decay of the disorder-averaged magnetization, $Z(t)$, as defined in the main text,
	is found to decay zero on a timescale that diverges exponentially in the system size.
}
  \label{fig:time-dependence}
\end{figure}

\paragraph{Numerical Analysis of $\Uf$.}
In order to confirm that resonances do not destroy TTSB, it behooves us to
numerically analyze Eq.~(\ref{eqn:soluble-U_f}). First,
we use the time-evolving block decimation (TEBD) scheme \cite{Vidal2004} to compute the time evolution
of the short-range correlated initial state $[\cos(\pi/8) |\uparrow\rangle + \sin(\pi/8) {|\downarrow\rangle}]^{\otimes L}$ for system size $L=200$ and $h=0.3$ and $t_0 = 1$.
The top panel of Figure~\ref{fig:time-dependence} shows the expectation values of the Pauli spin operators, averaged over 146 disorder configurations and over the spatial interval $i \in [50,150]$.
The TEBD calculations were done with Trotter step $0.01T$ and bond dimension $\chi = 50$.
The spin-flip part of the Floquet operator is applied instantaneously, which explains why the oscillation appears to be step-like. After an initial transient, the expectation values oscillate at frequency $\pi/T$, half the drive frequency.

Lest a skeptic wonder whether such oscillations continue to much later times of decay just beyond the times accessible by TEBD,
we analyze smaller systems by numerical exact diagonalization (ED) of the Floquet operator.
To extract the time on which the magnetization decays, we consider the time evolution of the magnetization starting from random initial product states
that are polarized in the $z$ direction, and compute the average $Z(t) = \overline{(-1)^t \langle \sigma^z_i(t) \rangle \mathrm{sign} (\langle \sigma^z_i(0) \rangle)}$
over 500 disorder realizations and for a fixed position $i$.
As shown in the bottom panel of Fig.~\ref{fig:time-dependence}, there is an initial decay of this quantity, which for the parameters chosen here occurs around
$t/T = 10$, and then a plateau that extends up to a time that diverges exponentially in the system size, and even for these small system sizes reaches
times comparable to the inverse floating point precision.
In the Supplementary Material, we explore these timesales in more detail and desribe ways in which signatures of TTSB can be observed for individual disorder configurations (without disorder averaging).

%We explore these timescales in more detail in the Supplementary Material.
%In individual disorder realizations, $\langle \sigma^x_i \rangle$ and
%$\langle \sigma^y_i \rangle$ are noisier. However, one can still observe a clear signature of TTSB by looking at the Fourier transform of the time dependence of a single disorder realization, as we discuss in the Supplementary Materials.

\begin{figure}
  \includegraphics{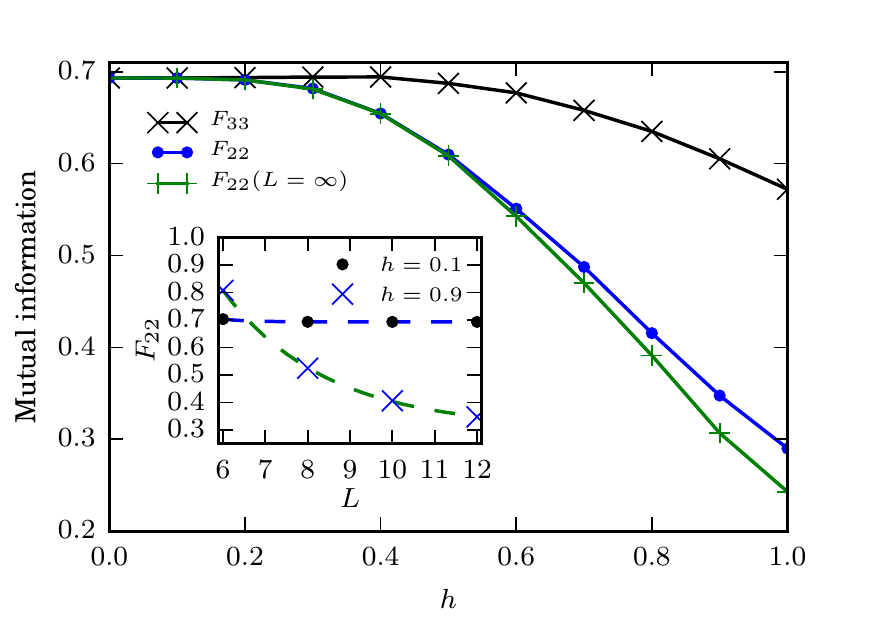}
  \caption{(Color online) The mutual information between the $n$ left- and rightmost sites, $F_{nn}$, for $n=2$ and $n=3$. The main panel
  shows results for $L=12$, as well as the extrapolated value of $F_{22}$ for $L \rightarrow \infty$. To extrapolate, we fit
  $F_{22}(L) = F_{22}(\infty) + c e^{-L/\xi}$, with $F_{22}(\infty)$, $c$ and $\xi$ fit parameters.
  Example fits for $h=0.1$ and $h=0.9$ are shown in the inset.  }
  \label{fig:mutual-information}
\end{figure}

We now turn to ED of the Floquet operator to verify that TTSB-2 holds.
We diagonalize $\Uf$ for $L=6, 8, 10, 12$ sites and 3200 disorder realizations and compute the mutual
information between the left- and rightmost $n$ sites, labelled $F_{nn}$.
We find that the mutual information obeys the scaling form:
$F_{nn}(h,L) = F_{nn}(g,\infty) + {c_n} \exp(-L/\xi(h))$. We expect that
$F_{nn}(h,\infty)=0$ in the TTS-invariant phase, $h>h_c$; and
$F_{nn}(g,\infty)>0$ in the TTSB phase, $h<h_c$, with 
$F_{nn}(g,\infty)\rightarrow \ln 2$ as $n\rightarrow \infty$.
The results in Fig. \ref{fig:mutual-information} are consistent with this form,
with  ${h_c}\stackrel{>}{\scriptstyle \sim} 1$.
It is remarkable that scaling holds even for such small systems, and that
$F_{22} \approx F_{33} \approx \log 2$ for $h < 0.3$; evidently, $L=12$ and $n=2, 3$
are not so far from the thermodynamic limit.

\paragraph{Implications of TTSB.}
In systems exhibiting MBL, it is commonly thought that there exists a \emph{complete} set of local integrals of motion (LIOMs): that is, there is a set of quasi-local operators $\tau_i^z$ which commute with each other and with the Floquet operator $\Uf$ (or the Hamiltonian in the static case), and such that the eigenvalues of $\tau_i^z$ uniquely specify a state in the Hilbert space \cite{Huse2013a,Serbyn13b,Huse14}. Systems with TTSB violate this principle. Indeed, in our model at its soluble point at $h=0$, the locally indistinguishable states
$(\exp(i t_0 {E^-}(\{s_i \}) /2) |\{s_i \}\rangle \pm \exp(-i t_0 {E^-}(\{s_i \}) /2)|\{-s_i \}\rangle)/\sqrt{2}$ are eigenstates, with different quasienergy.
No LIOM can distinguish between these two states, so no set of LIOMs can be
complete. (Though the existence of a complete set of LIOMs is sometimes taken as
the \emph{definition} of MBL, the TTSB phase is still MBL in the sense of, for
example, long-time dynamics, since $(\Uf)^2$ does have a complete set of LIOMs).
By a similar argument, one can show that there does not exist a quasi-local effective Hamiltonian
 $H_{\mathrm{eff}}$ such that 
 $\Uf = \exp(-iTH_{\mathrm{eff}})$, whereas for Floquet-MBL systems without TTSB this is likely to be the case \cite{Abanin2014,Ponte15b}.

As noted earlier, the oscillations arise from the occurrence of multiplets of
states separated in Floquet eigenvalue by $\Omega/n$, where $\Omega = 2\pi/T$ is
the drive frequency.
We don't use this to identify the TTSB phase in ED because the states are too closely spaced in energy to pick
out such multiplets. However, their existence suggests that the system can
radiate at frequency $\Omega/n$. The fact that systems oscillating in time can
radiate has been cited as an argument against the existence of TTSB
\cite{Bruno13a,Bruno2013},
since a system maintaining persistent oscillations while simultaneously
radiating would be inconsistent with conservation of
energy. However, in the Floquet case, this is not an issue since energy is being
continually supplied by the drive. Nor does such persistent radiation violate conservation of
quasienergy, due the fact that physical (i.e.\ short-range correlated) states
are not quasienergy eigenstates.
%and in fact a process in
%which a system in such a
%state emits a photon of frequency $\Omega/n$, while the system itself remains
%unchanged, is perfectly consistent with quasienergy conservation.
(For details, see the Supplementary Material.)
On the other hand, in a system that breaks continuous TTS, radiation would cause the system
to decay to the ground state, which is reason to doubt that continuous TTSB can occur.

\paragraph{Discussion.}
The model Eqs. (\ref{eqn:soluble-U_f}) and (\ref{eqn:H_MBL}) is soluble at $h=0$ because the operator
$\exp(i \frac{\pi}{2} {\sum_i} {\sigma^x_i}) = {\prod_i} i{\sigma^x_i}$ that is applied
at the beginning of each driving cycle maps eigenstates of
$H_\text{MBL}$ to eigenstates of $H_\text{MBL}$. Analogous soluble models can be constructed for $\mathbb{Z}_n$ spins
in which time translation by $T$ is broken down to $nT$.

Our model has no symmetries, other than discrete time-translation
symmetry. Hence, the $\ln 2$ that we find in the mutual information must be a consequence of TTSB; there is no other symmetry
to break. However, TTSB can occur in models with other symmetries. A particularly interesting example is given
by symmetry-protected topological (SPT) phases of Floquet-MBL systems~\cite{vonKeyserlingk16a,Else16a,Potter16,Roy16}.
In $d$-dimensions, such phases are classified by $H^{d+1}(G\times \mathbb{Z},U(1))=H^{d+1}(G,U(1))\times H^{d}(G,U(1))$
\cite{Else16a}. The second factor on the right-hand-side of this equality is a $(d-1)$-dimensional SPT phase that is
`pumped' to the boundary with each application of the Floquet operator, thereby
breaking TTS on the boundary.
%Meanwhile, the models of Refs.~\cite{Khemani2015a,vonKeyserlingk16b} break another symmetry spontaneously in addition to TTSB
%(though the TTSB was not identified as such).
%However, it is worth re-emphasizing, additional symmetries (whether preserved or spontaneously-broken)
%are not essential to TTSB and are not present in our model.

The definition TTSB-1 naturally suggests an experiment that could observe the phenomenon predicted here.
Signatures of MBL have been observed in trapped systems of neutral atoms~\cite{Schreiber15} and trapped ions~\cite{Smith15},
and signatures of single-particle localization have been seen in coupled superconducting qubits 
\cite{Chen14}.
%In the model given by Eqs. (\ref{eqn:soluble-U_f}) and (\ref{eqn:H_MBL}),
%we used a longitudinal field $h^z_i \sigma^z_i$ to break the integrability
%of the system, and this is a natural term in the Hamiltonian of coupled superconducting qubits. For systems of
%ultra-cold fermions on a lattice this term is highly non-local after Jordan-Wigner transformation, but another term,
%such as $J^x_i \sigma^x_i \sigma^x_j$ could serve the same role.
In any of these systems, one can prepare an arbitrary initial product state,
evolve to late times according to a drive in the class considered here, and measure the ``spins'' in the desired basis.
Our prediction is that persistent oscillations will be observed at a fraction of the drive frequency.

\begin{acknowledgments}
D.E. acknowledges support from the Microsoft Corporation.
\end{acknowledgments}

%\bibliography{topo-phases,../References/references-processed}
%\putbib[../References/references-processed,../References/references-extra]

\end{bibunit}

\clearpage

\begin{bibunit}[apsrevmod]
\section{Supplementary Materials}
\newcommand{\EqOneRef}{Eq.~(\ref{eqn:soluble-U_f})}

%\renewcommand*{\citenumfont}[1]{S#1}
%\renewcommand*{\bibnumfmt}[1]{[S#1]

%%%%%%%%%%%%%%%%%%%%%%%%%%%%%%%%%%%%%%%%%%%%%%%%%%%%%%%%%%%%%%%%%%%%%%%%%%%%%
%\renewcommand{\thesection}{S.\arabic{section}}
%\renewcommand{\thesubsection}{\thesection.\arabic{subsection}}
\setcounter{equation}{0}
\setcounter{figure}{0}
% Hack for making SOM Equations Conform to Science Format
%
% e.g. (S1), (S2), etc
% Hack for making SOM Equations and Figures conform to Science Format % % e.g. (S1), (S2), etc

\renewcommand{\theequation}{S\arabic{equation}}

% Hack for making figures Say \figurename S\thefigure, e.g. Figure S1:
%\makeatletter
%\makeatletter
%\renewcommand{\fnum@figure}{\figurename~S\thefigure}
%\makeatother

\renewcommand{\thefigure}{S\arabic{figure}}

% use bibnumfmt to change style at the end of the document
\renewcommand{\bibnumfmt}[1]{[S#1]}
% citenumfont command adds S to all numbers
\renewcommand{\citenumfont}[1]{S#1}
 
\renewcommand{\figurename}{Figure}
%%%%%%%%%%%%%%%%%%%%%%%%%%%%%%%%%%%%%%%%%%%%%%%%%%%%%%%%%%%%%%%%%%%%%%%%%%%%%

\subsection{Local structure of Floquet perturbation theory}
\label{sec:perturbation_theory}

Consider a soluble Floquet operator
\begin{equation}
U_\text{f}^0 = \mathcal{T} e^{-i\int_0^1 H_0(t) dt},
\end{equation}
and a time-dependent local perturbation $\lambda V(t)$, and define
\begin{equation}
    U_\text{f} = \mathcal{T} \exp\left(-i\int_0^1 [H_0(t) + \lambda V(t)] dt\right), \quad
    \mathcal{T} = \mbox{time-ordering}.
\end{equation}
By Trotterizing, we can show that
\begin{equation}
U_\text{f}  = U_\text{f}^0  \times \mathcal{T} \exp\left(-i\int_0^1 (U_f^0)^{\dagger}(t) \lambda V(t) {U_f^0}(t)\right),
\end{equation}
where ${U_0}(t) = \mathcal{T} e^{-i\int_0^{t} H_0(t^{\prime})}$. Hence, without loss of generality we can just consider a perturbed Floquet operator $U_\text{f} = U^0_\text{f} \, U'$, where 
where $U' = \mathcal{T} \exp\left(-i\int_0^1 \lambda V(t) dt\right)$ for some local time-dependent $V$.
We label the eigenstates of $U^0_\text{f}$ as $\ket{i}$, with quasienergies $\omega_i$.
We will now construct, order-by-order, a local unitary rotation that diagonalizes the perturbation.

\paragraph{First order.} At first order we look for a unitary $e^{i \lambda S}$ such that $e^{i\lambda S} U_f e^{-i\lambda S}$ is diagonal (to first order in $\lambda$).
Expanding $e^{i \lambda S} U^0_\text{f} U' e^{-i \lambda S}$ to first-order in $\lambda$ and taking the matrix elements with $\bra{i}$ and $\ket{j}$, we see that
we can make it diagonal to this order
by taking:
\begin{equation}
\label{Sfirst}
\bra{i} S \ket{j} = \frac{\bra{i} \overline{V} \ket{j}}{e^{i(\omega_i-\omega_j)} - 1}\, \quad (i \neq j)
\end{equation}
where $\overline{V} = \int_0^1 V(s) ds$. We can choose to set
$\bra{i} S \ket{i} = 0$.

It might not be clear whether this $S$ is local, given that the eigenstates $\ket{i}$ might be highly non-local ``cat states''. To see that it is, we adapt an idea originally due to Hastings~\cite{Hastings2003} (as refined in Ref.~\cite{Osborne2007}) to the Floquet case. First write $\overline{V}$ as a sum of local terms $\overline{V} = \sum_X \overline{V}_X$, where $\overline{V}_X$ is supported on a bounded region $X$. Then we can write $S = \sum_X S_X$, where
\begin{equation}
S_X = \sum_{i \neq j} \frac{\ket{i}\bra{i} \overline{V}_X\ket{j} \bra{j}}{e^{i(\omega_i - \omega_j)} - 1} \equiv \sum_{i \neq j} f(\omega_i - \omega_j) \ket{i} \bra{i} \overline{V}_X \ket{j} \bra{j}, \label{SX}
\end{equation}
Now suppose that there are no ``resonances'' near $X$, by which we mean that $|e^{i(\omega_i - \omega_j)} - 1| > \gamma > 0$ for all $i,j$ for which the matrix element $\bra{i} \overline{V}_X \ket{j}$ is nonzero. Then we can replace $f(\omega)$ with $\widetilde{f}(\omega)$ in Eq.~(\ref{SX}), where $\widetilde{f}(\omega)$ is an infinitely differentiable function with period $2\pi$ such that $\widetilde{f}(0) = 0$ and $\widetilde{f}(\omega) = f(\omega)$ for $|e^{i(\omega_i - \omega_j)} - 1| > \gamma$. By taking matrix elements one can then verify that
\begin{equation}
    S_X = \sum_{n = -\infty}^{\infty} a_n (U^0_\text{f})^{-n} \overline{V}_X (U^0_\text{f})^n,
\end{equation} where $a_n$ are the Fourier series coefficients of $\widetilde{f}$:
$\widetilde{f}(\omega) = \sum_{n = -\infty}^{\infty} e^{in\omega} a_n$.
From this, we can show that $S_X$ is quasi-local provided that $U^0_\text{f}$ obeys a Lieb-Robinson bound. In particular, however, if we choose $U^0_\text{f}$ such that $(U^0_\text{f})^n$ doesn't increase the support of operators by more than an $n$-independent constant,
we see that $S_X$ is still strictly local on a region of slightly larger size. In particular, this can be shown to be true of the Floquet operator
$U^0_\text{f}$ in \EqOneRef{} in the main text.
To see this, note that in $(U^0_\text{f})^n$ we can move all the spin flips to the end at the cost of simply changing the sign of the $h_i$'s during the course of the evolution, and the time evolution of a Hamiltonian which is the sum of terms, each of which is a product of Pauli $\sigma_z$ operators (even if the coefficients vary with time) never increases the support of operators by more than a constant amount.

\paragraph{All orders.} Suppose that we have found a unitary rotation which diagonalizes the perturbation to order
$\lambda^n$, such that
$U_\text{f} = U^0_\text{f} U'$, with
\begin{equation}
\cV = \exp\left( -i \left\{V_d + \lambda^{n+1} V_{nd} + O(\lambda^{n+2}) \right\} \right),
\end{equation}
where $V_{d}$ is diagonal, $V_{d} = O(\lambda)$ and $V_{nd}$ is non-diagonal. (At first-order, i.e.\ $n = 0$, if $U'$ was originally the evolution of a time-dependent Hamiltonian we can still generate such an expression for $\cV$ using the Campbell-Baker-Haussdorf formula.) Then we want to find $S$ such that $e^{iS} U_\text{f} e^{-iS}$ is diagonal to order $\lambda^{n+1}$, or equivalently, writing $e^{iS} U_\text{f} e^{-iS} = U^0_\text{f} U'^{\prime}$, that $U'^{\prime}$ is diagonal. We see that
\begin{equation}
U'^{\prime} = U_0^{\dagger} e^{iS} U' e^{-iS} = e^{i U_0^{\dagger} S U_0} U' e^{-iS}.
\end{equation}
From the Campbell-Baker-Haussdorf formula, we see that
\begin{equation}
\label{cbh}
U'^{\prime} = \exp\left(i\left\{ -V_d - \lambda^{n+1} V_{nd} + U^0_\text{f} S (U^0_\text{f})^{\dagger} - S \right\} + O(\lambda^{n+2})\right),
\end{equation}
and hence we set the expression in $\{ ... \}$ to be diagonal. Taking off-diagonal matrix elements gives
\begin{equation}
\label{Sn}
\bra{i} S \ket{j} = \frac{\bra{i} \lambda^{n+1} V_{nd} \ket{j}}{e^{i (\omega_i - \omega_j)} - 1} \quad (i \neq j),
\end{equation}
and we choose to set $\bra{i} S \ket{i} = 0$.
We can then repeat the process at next order, with $n^{\prime} = n+1$, $V_d^{\prime} = V_d + \lambda^{n+1}V_{nd} - (U^0_\text{f})^{\dagger} S U^0_\text{f} + S$, and $V_{nd}^{\prime}$ equal to the coefficient of $\lambda^{n+2}$ in the Campbell-Baker-Haussdorf expansion  Eq.~(\ref{cbh}).

We observe that at all orders in the perturbation theory, locality is preserved. The only operations contained in the exponentials are addition, conjugation by $U^0_\text{f}$, taking nested commutators (through the Campbell-Baker-Haussdorf expansion), and evaluating expressions of the form \eqnref{Sn}. The first three \emph{manifestly} preserve locality, and the last one preserves locality in the absence of resonances for the same reasons discussed in the first-order section above. Therefore, the unitary rotation that relates the eigenstates of $U_\text{f}$ to the eigenstates of $U^0_\text{f}$ is indeed a \emph{local} unitary at all orders.

\subsection{Numerical Observation of Persistent Oscillations at Very Late Times}
 
\begin{figure}
\includegraphics[width=8cm]{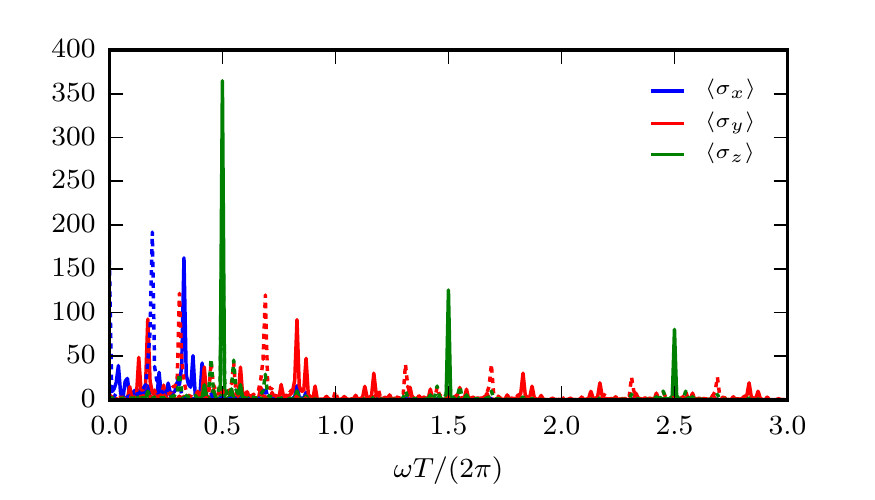}
  \caption{The Fourier transform of the time evolution at late times (taken over the interval $200<t<300$)
for two individual disorder realizations (shown as solid and dotted lines
respectively), at $h=0.3$. The dominant peaks at $\omega = (2k+1)\pi/T$ are universal, whereas the smaller
peaks at other locations are disorder-dependent.}
  \label{fig:single--disorder realization}
\end{figure}

\begin{figure}
  \includegraphics{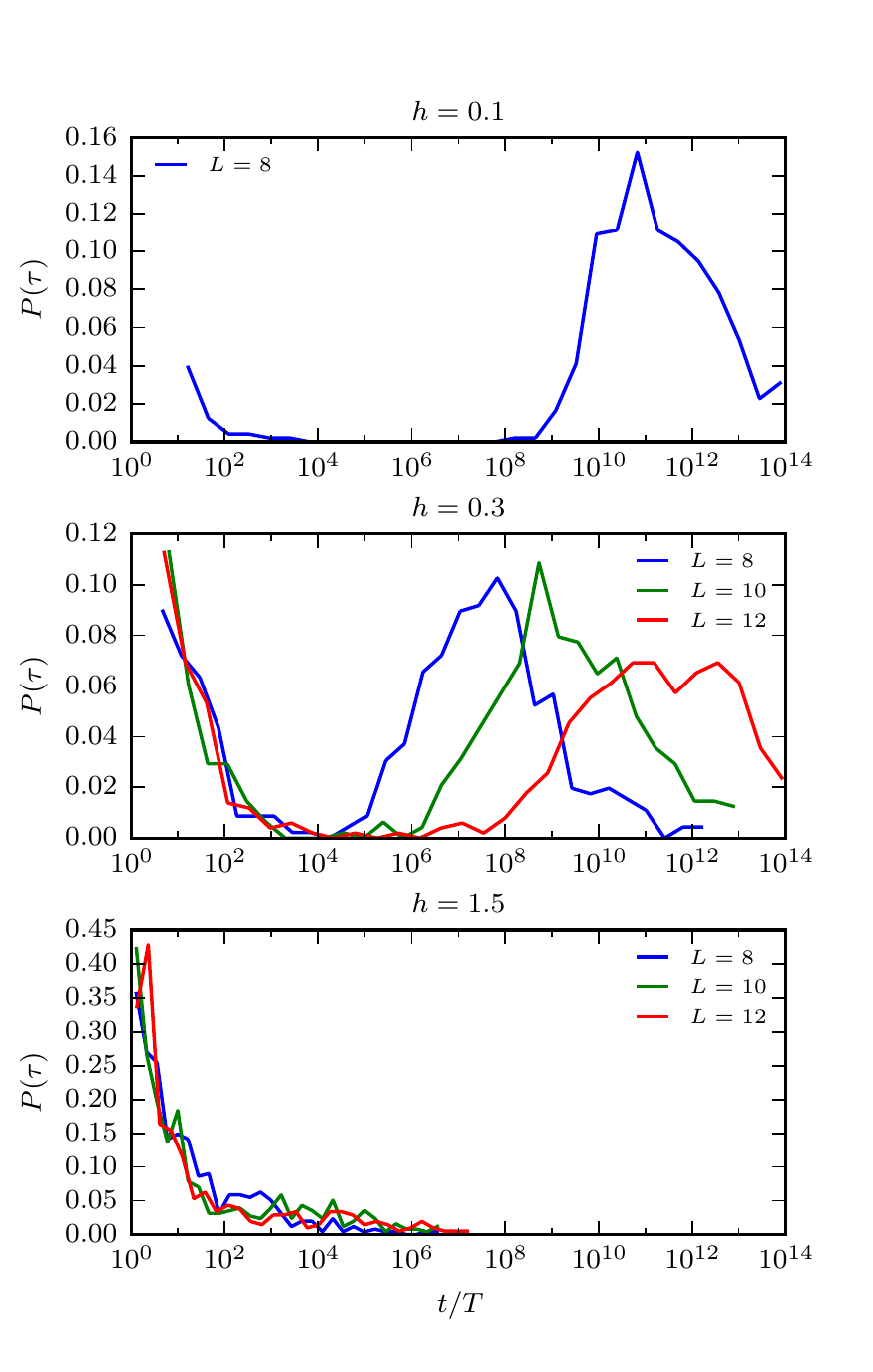}
  \caption{Histogram of the characteristic timescale $\tau$, as defined in the text, for different values of
   the magnetic field. From top to bottom: $h=0.1$ deep in the TTSB phase, $h=0.3$ in the same regime as discussed in the
  main text, and $h=1.5$ beyond the TTSB phase. \label{fig:Ptau} }
\end{figure}

In a single disorder realization, we can go to much later times in TEBD. Moreover,
experiments might be carried out in a small number of disorder realizations.
As noted in the main text, $\langle \sigma^x_i \rangle$ and
$\langle \sigma^y_i \rangle$ are noisier in individual disorder realizations.
However, one can still observe a clear signature of TTSB by looking at the Fourier transform of the time dependence of a single disorder realization, as shown in Fig. \ref{fig:single--disorder realization}. There are strong peaks at $\pi/T$, with subleading peaks at $(2k+1)\pi/T$, indicating the fractional frequency response. (The other peaks in the Fourier transform have their origins in the discreteness of the local quasienergy spectrum near a given point, and can be distinguished by the fact that their positions vary depending on the disorder realization.)
These results indicate that the oscillations persist to later times than shown in the upper panel of
Fig. \ref{fig:time-dependence} in the main text and that they are visible even in a single disorder realization.

To more carefully examine the decay of the oscillations, we turn again to exact diagonalization. For a given
disorder realization and initial state, we can determine a characteristic timescale $\tau$ by computing
$(-1)^t \langle \sigma^z_i(t) \rangle \text{sign}(\langle \sigma^z_i(0) \rangle)$. This is defined
such that it is positive for small times, and we define $\tau$ to be the time at which this observable
first changes its sign.
In Fig.~\ref{fig:Ptau}, we show a histogram of these $\tau$ for different system sizes and strengths
of the magnetic field. We observe a very interesting structure: deep in the TTSB phase, at
$h=0.1$, we find a single large peak at very large times (here, we show only $L=8$ since for
larger systems the $\tau$ are too large compared to the floating point precision). In an intermediate
range, such as $h=0.3$, we find two pronounced peaks, where the location of the first peak does not depend
on system size while the second peak is centered around a time that diverges exponentially. The relative
weight of the two peaks seems unaffected by system size. In this regime, the average of $\tau$ is dominated
by rare instances with very large $\tau$, while the typical value is dominated by instances with short 
characteristic times. In the disorder-averaged value of the magnetization $Z(t)$, which was discussed in
the main manuscript, the first peak in the distribution of $\tau$ manifests in the decay from the initial
value to the intermediate plateau, and the second peak corresponds to the decay from this plateau to zero.
Finally, in the limit of very large $h$ where the system has been driven out of the TTSB phase, we find
the histogram to be dominated by a peak at short times.
\subsection{Radiation Emitted from a TTSB System at Lowest Order in Perturbation Theory}

Let us suppose, for illustrative puposes, that our spin system is coupled to the electromagnetic field through
the Jaynes-Cummings Hamiltonian
\begin{equation}
{H_1} = V (a + {a^\dagger}), \quad V = g \sum_i \sigma^z_i,
\end{equation}
where $a^\dagger$, $a$ are creation/annihilation operators for photons of frequency $\Omega/2$ (where $\Omega = 2\pi/T$ is the drive frequency.)
This operator ${\sigma^x_i}$ conventionally appears in this Hamiltonian; our Hamiltonian is rotated in spin
space relative to the conventional one. The transition amplitude between initial and final Floquet eigenstates
$| m\rangle$, $| n\rangle$ is given, in the interaction picture, by:
\begin{equation}
A_{m,n} = \langle n,1 | \, {\cal T} \exp\left(-i \int_{-\infty}^{\infty} dt {H_1}(t^{\prime})\right) \, | m,0 \rangle
\end{equation}
where $\langle n,1|$ is the state with the spin system in the state $\ket{n}$ and a single photon (and similarly for $\ket{m,0}$),
and ${H_1}(t) \equiv {U_0^\dagger}(t,-\infty) {H_1} {U_0}(t,-\infty)$ and ${U_0}(t,-\infty) \equiv {\cal T} \exp\bigl(-i \int_{-\infty}^t dt' {H}(t')\bigr)$. The unperturbed Hamiltonian $H(t)$ is the stroboscopic Hamiltonian given in \EqOneRef{} in the main text and the text below it. We write $t = kT + s$, where $s\in [0,T)$. Then we can write
${U_0}(t,jT) = {U_0}(s,0) U_\text{f}^{k-j}$. To lowest-order, the transition amplitude can be written in the form:
\begin{widetext}
\begin{align}
A_{m,n} &= -i \sum_{k=-\infty}^\infty \int_0^1 ds \langle n,1 |
(U_\text{f}^{k})^\dagger {U_0^\dagger}(s,0) {H_1} {U_0}(s,0) U_\text{f}^{k}
 | m,0 \rangle\cr
&= -i \sum_{k=-\infty}^\infty e^{ikT (\omega_n - \omega_m + \Omega/2)} \int_0^1 ds \langle n|
 {U_0^\dagger}(s,0) V {U_0}(s,0) | m \rangle \\
&\equiv -i f(\omega_n - \omega_m) \int_0^1 ds \langle n|
 {U_0^\dagger}(s,0) V {U_0}(s,0) | m \rangle \label{general_amplitude}
\end{align},
\end{widetext}
where
\begin{equation}
f(\omega) = \frac{2\pi}{T} \sum_{k = -\infty}^{\infty} \delta\left(\omega + \left[k+\frac{1}{2}\right]\Omega\right).
\end{equation}

This matrix element is generally non-zero. For instance, consider the soluble point $h=0$, where we take the spin-flip part of the Floquet operator to act instantaneously such that $\int_0^1 ds U_0^{\dagger}(s,0) \sigma_i^z U_0(s,0) = \sigma_i^z$.
Then for initial and final states are $|\pm\rangle \equiv (\exp(i t_0 {E^-}(\{s_i \}) /2) |\{s_i \}\rangle \pm \exp(-i t_0 {E^-}(\{s_i \}) /2)|\{-s_i \}\rangle)/\sqrt{2}$, we find that $\langle - | {\sigma^z_i} | + \rangle = \langle + | {\sigma^z_i} | - \rangle = 1$ for any $i$, and hence 
\begin{equation}
\label{transition_amplitude}
A_{+-} = A_{-+} = -\frac{2 \pi i gN}{T} \delta(0).
\end{equation}
Now consider a locally-prepared initial state, such as 
\begin{equation}
\label{initstate}
|\{s_i \}\rangle = (|+\rangle \,\pm \,|-\rangle)/\sqrt{2},
\end{equation}
[Here we have set $h^z_i = 0$ in order to unclutter the equations, so that $E^{-}(\{s_i\}) = 0$.] Then, in the absence of a coupling to the electromagnetic field, it would not change with time in the interaction picture. (The fractional frequency response in the interaction picture comes from the time evolution of observables.) 
However, \eqnref{transition_amplitude} tells us that at lowest-order in perturbation theory, the system can emit a photon at frequency $\Omega/2$ and transition from $\ket{-} \leftrightarrow \ket{+}$. However, this only changes the superposition \eqnref{initstate} by a global phase factor $\pm 1$.
One might wonder why this does not violate conservation of quasienergy, given that a photon of frequency $\Omega/2$ has been emitted. However, we observe that the state \eqnref{initstate} is not a quasienergy eigenstate; rather, it is a superposition of two eigenstates with quasienergies differing by $\Omega/2$. Therefore, its quasienergy is only well-defined modulo $\Omega/2$. We note that neither $|+\rangle$ nor $|-\rangle$ is ``higher'' in quasienergy.
The system can emit a photon of energy $\Omega/2$ while transitioning from $|+\rangle$ to $|-\rangle$ or from $|-\rangle$ to $|+\rangle$
since  since $-\Omega/2 \equiv \Omega/2 \pmod{\Omega}$. [Mathematically, this corresponds to the statement that $f(\Omega/2) = f(-\Omega/2)$ in \eqnref{general_amplitude}.]

%\putbib[../References/references-extra,../References/references-processed]

\end{bibunit}

\end{document}